\documentclass[]{article}
\usepackage{graphicx}
\usepackage{amsmath, subfigure, latexsym}
\usepackage{fullpage, comment}
\usepackage{algorithm}
\usepackage{algpseudocode}
\usepackage{hyperref}

\newtheorem{conjecture}{Conjecture}
\newtheorem{theorem}{Theorem}
\newtheorem{claim}{Claim}
\newtheorem{definition}{Definition}
\newtheorem{ex}{Example}
\newtheorem{prop}{Proposition}

\title{On generating $k$-factorable graphic sequences with connected (resp. no connected) $k$-factors}
\author{Asish Mukhopadhyay, Daniel John and Lucas Sarweh\\
School of Computer Science \\
Univesity of Windsor\\
Ontario, Canada}

\begin{document}

\maketitle

\begin{abstract}
In this note, we consider the problem of generating $k$-factorable graphic sequences with  connected (resp. no connected) $k$-factors.
\end{abstract}

\section{Introduction}
Let $d = \langle d_1, d_2, \ldots, d_n \rangle$ be a finite sequence of integers such that $d_1 \geq d_2 \geq \ldots \geq d_n > 0$.

\begin{definition}
If there exists a graph $G = (V, E)$ with vertices $v_1, v_2, \ldots, v_n$ such
that the degree, $d(v_i)$, of vertex $v_i$ is $d_i$, for $i = 1, \ldots, n$ then the sequence $d$ is said to be graphic and $G$ is said to be a realization of $d$.
\end{definition}

Erd\H{o}s and Gallai \cite{erdosGallai1960} showed that:

\begin{theorem}
	A sequence $\langle d_1 ,\ldots , d_n \rangle$ of nonnegative integers in nonincreasing order is graphic if and only if:
	
	\textnormal{(a)} $\Sigma_{i=1}^n d_i$ even and,\\
	\textnormal{(b)} for each integer $k$ with, $1 \leq k \leq n$,
	$$
	\Sigma_{i=1}^k d_i \leq k(k-1) + \Sigma_{i=k+1}^n~ \textrm{min} ~\{k , d_i \}
	$$
	holds.
\end{theorem}

These are the well-known Erd\H{o}s and Gallai Inequalities, EGI for short.

\begin{definition}
A graphic degree sequence $\langle d_1, d_2, \ldots, d_n \rangle$ is said to be $k$-factorable if there exists a 
graphic realization that has a $k$-regular spanning graph as a subgraph. 
\end{definition}

{\bf Note:} A more general version of the problem, known as the $f$-factor problem, is this: Given 
an  undirected graph $G = (V, E)$ and a map $f : V  \rightarrow N \cup \{0\}$, the $f$-factor theorem of Tutte~\cite{tutte_1952} gives necessary and sufficient 
conditions under which there exists a subgraph $H$ of $G$ that realizes the vertex degrees, given by the map $f$. This problem and its variations have been extensively
researched (see \cite{Akiyama1985FactorsAF}, \cite{DBLP:journals/gc/KouiderV05}).\\

Rao and Rao~\cite{RAMACHANDRARAO1972185} made the following conjecture:

\begin{conjecture}
A graphic degree sequence $(d_1, d_2, \ldots, d_n)$ is $k$-factorable if and only
if the sequence $(d_1 - k, d_2 - k, \ldots, d_n - k)$ is graphic.	
\end{conjecture}

Kundu~\cite{KUNDU1973367} proved the slightly more general version of this conjecture:

\begin{theorem}
Let $(d_1, d_2, \ldots, d_n)$ and $(d_1 - k, d_2 - k, \ldots, d_n - k)$ be two graphical sequences with the
property that for some $k \geq 0$, $k \leq k_i \leq k + 1$ for all $i$. Then there exists
a graph with degree sequence $(d_1, d_2, \ldots, d_n)$, containing a $(k_1, k_2, \ldots, k_n)$-factor.	
\end{theorem}

In this note, we consider the problem of generating $k$-factorable graphic  sequences that have connected (resp. no connected) $k$-factors. Towards this,
as a first step, we address the problem of generating graphic sequences. \\

We can try to solve the EGI inequalities to determine a sequence $d$ that satisfies these. Unfortunately, the 
\textit{min}-function on the right-hand side of these inequalities complicates the task. What comes to our rescue are a set of sufficient conditions in \cite{DBLP:journals/dm/ZverovichZ92} that make no direct reference to the EGIs.\\

Choose 
integer parameters $a$ and $b$ such that $a \geq b > 0$, and let $K(a, b)$ denote the class of sequences
$d$ such that: 
\begin{equation} \label{seqChoice}
a \geq d_1 \geq d_2 \geq \ldots \geq d_n \geq b > 0
\end{equation}

The following result shows that each sequence in the class $K(a, b)$ is graphic. Let $l = (a + b + 1)^2/4b$.

\begin{theorem}\label{zzThm}\textnormal{\cite{DBLP:journals/dm/ZverovichZ92}}
 If $d \in K(a, b)$, such that $\Sigma_{i=1}^n d_i$ is even and $n \geq l$, then $d$ is graphic.
\end{theorem}

An intuitive interpretation of Theorem~\ref{zzThm} is that if we choose a sequence $d$ much longer than $l$, 
then $d$ is assuredly graphic. This result has been extended and generalized in \cite{DBLP:journals/dm/BarrusHJW12}.\\

In the next two sections we use this result to propose algorithms for generating $k$-factorable graphic sequences with (a) connected $k$-factors (b) no connected $k$-factors.

\section{Generating $k$-factorable graphic sequences with connected $k$-factors}

In \cite{RAMACHANDRARAO1972185} it is shown that a graphical degree sequence $d = \langle d_1, d_2, \ldots, d_n \rangle$ with a $k$-factor has a connected $k$-factor iff the following condition holds for $s < n/2$.

\begin{equation}\label{raoInequalities}
\Sigma_{i=1}^{s} d_i < s(n-s-1) + \Sigma_{i=0}^{s-1} d_{n-i}
\end{equation}

The 2-factorable graphic sequence $dSix = \langle 3, 3, 2, 2, 2, 2 \rangle$ is easily seen to satisfy the above inequalities and thus there 
exists a graphic realization with a connected 2-factor. Fig.~\ref{picdSix} below shows a graphic realization of $dSix$ and a connected 2-factor.

\begin{figure}[h!]
	\centering
	\includegraphics[scale =0.6]{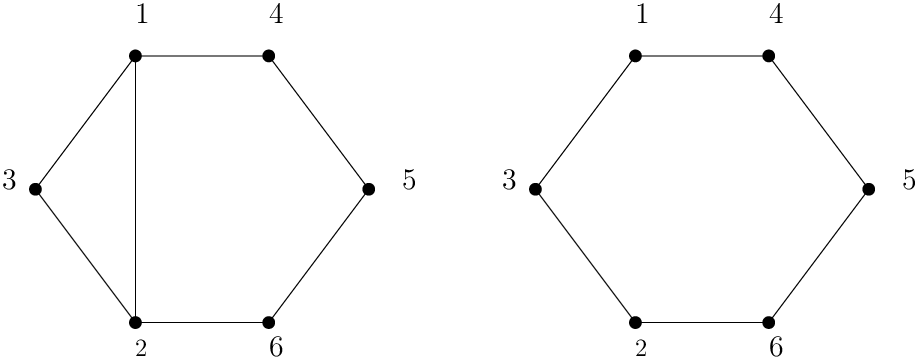}
	\caption{\em{A graph on dSix with a connected 2-factor}}
	\label{picdSix}
\end{figure}

The following trial-and-error heuristic can be used to generate a $k$-factorable graphic sequence with a connected $k$-factor. \\

\begin{algorithm}
	\caption{Heuristic for generating $k$-factorable graphic sequence with a connected $k$-factor}
	\begin{algorithmic}[1]
		\State \textbf{Input:} $n, k, l$
		\State \textbf{Output:} A $k$-factorable graphic sequence
		\Procedure{$k$-factorable graphic sequence}{$n, k, l$}
		  \State Choose parameters $a$ and $b$ such that $a \geq b > 0$ and an integer $k \geq 2$.
		  \State Choose a positive integer sequence satisfying Theorem 3.
		  \State Verify that $d-k$ is also graphic. If not, pick another sequence
		  \State If both $d$ and $d-k$ are graphic check if all the inequalties in Eqn.~\ref{raoInequalities} are satisfied
		  \State If so, return this sequence, else go to to Step 5.
	  \EndProcedure
\end{algorithmic}
\end{algorithm}

\begin{ex}
For $a = 10$ and $b = 3$, $n \geq 17$; set $k = 3$. \\

The following sequence $d = (10, 10, 10, 10, 9, 9, 9, 9, 8, 8, 8 , 8, 7, 7, 7, 7, 6, 4)$
has a connected 3-factor as seen in a realization in Fig.~\ref{fig02}. 

\begin{figure}[h!]
	\centering
	\includegraphics[scale =0.4]{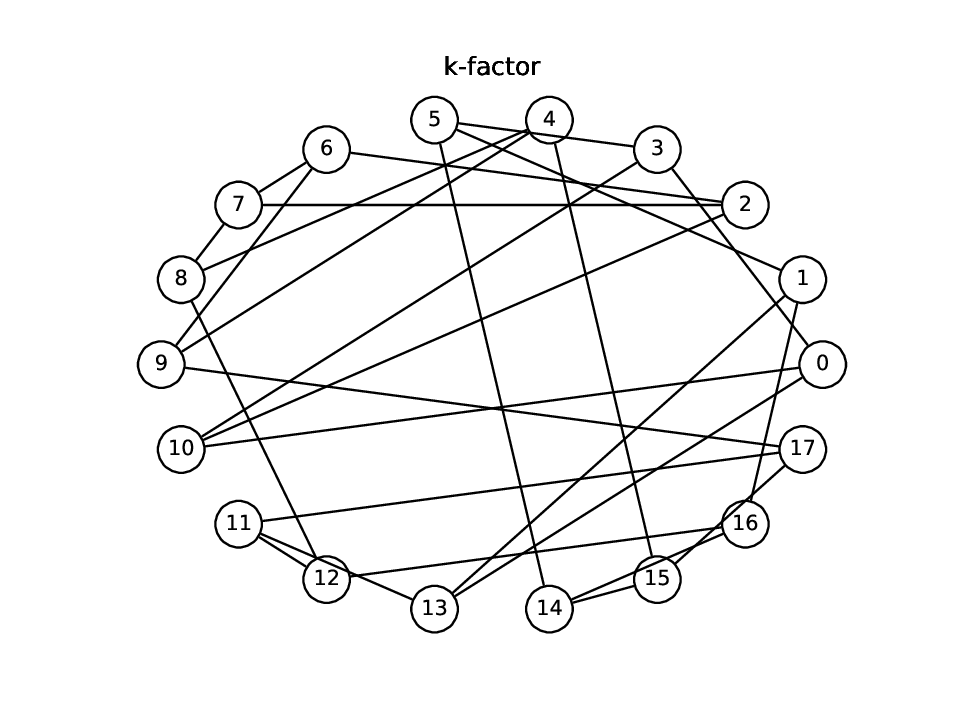}
	\caption{\em{Connected 3-factor for the degree sequence (10, 10, 10, 10, 9, 9, 9, 9, 8, 8, 8 , 8, 7, 7, 7, 7, 6, 4)}}
	\label{fig02}
\end{figure}
\end{ex}

\subsection{Further considerations}
Can the trial-and-error element be removed from the above heuristic for generating a $k$-factorable degree sequence with a connected $k$-factor ?\\

We note that the inequalities in Eqn.(\ref{raoInequalities}) are all satisfied if $(a-b)*n/2 < n-2$. This is obtained by taking the 
maximum number of terms in the sum $\Sigma_{i=1}^s(d_i - d_{n-i})$ and maximizing each summand by replacing it with 
$(a-b)$. \\

At the same time, we minimize the first term on the right-hand side of the inequalities by letting $s=1$ to give us $n-2$. Thus if $n/2*(a-b) < n-2$, then all inequalities are satisfied. We now choose the length of the sequence $d$, to be $n > max \{4/(2+b-a), (a+b+1)^2/4b\}$, assuming $a-b \neq 2$. \\

Below, we propose an algorithm to generate $k$-factorable graphic sequences.
\begin{algorithm}
	\caption{Algorithm for generating $k$-factorable graphic sequences with a connected $k$-factor}
	\begin{algorithmic}[1]
		\State \textbf{Input:} $a$, $b$ : Where $a\geq b > 0$, $2\neq a - b$
		\State \textbf{Output:} A $k$-factorable graphic sequence
		\Procedure{generateSequenceConnected}{$a, b$} 
			\State Find a large enough value of $n$ using $n = l + 1 = max(4/(2+b-a), ((a+b+1)^2)/4b) + 1$
			\State $sum = 0$
			\For {$i = 1$ to $n-1$}
				\State Generate a random number $x$, where $0 < b\leq x\leq a$ and append it to the sequence
				\State $sum = sum + x$
			\EndFor
			\If{$sum$ is even}
				\If{$b$ is even}
					\State append $b$ to the sequence
				\Else
					\State append $b + 1$ to the sequence
				\EndIf
			\Else
				\If{$b$ is even}
					\State append $b + 1$ to the sequence
				\Else
					\State append $b$ to the sequence
				\EndIf
			\EndIf
			\State Sort the obtained sequence in non-increasing order
			\State return the sequence
		\EndProcedure
	\end{algorithmic}
\end{algorithm}

\begin{ex}
Fig.~\ref{fig03} shows a graphic realization of the sequence $d = (6, 6, 6, 6, 5, 5, 5, 5)$ and a connected 2-factor. Its length 8 has been chosen to satisfy the bound determined by plugging in $a = 6$ and $b = 5$.\\

 Note that this sequence also has a realization with a disconnected $k$-factor as seen in Fig.\ref{fig04}.\\

\begin{figure}
	\centering
	\subfigure{\includegraphics[scale =0.4]{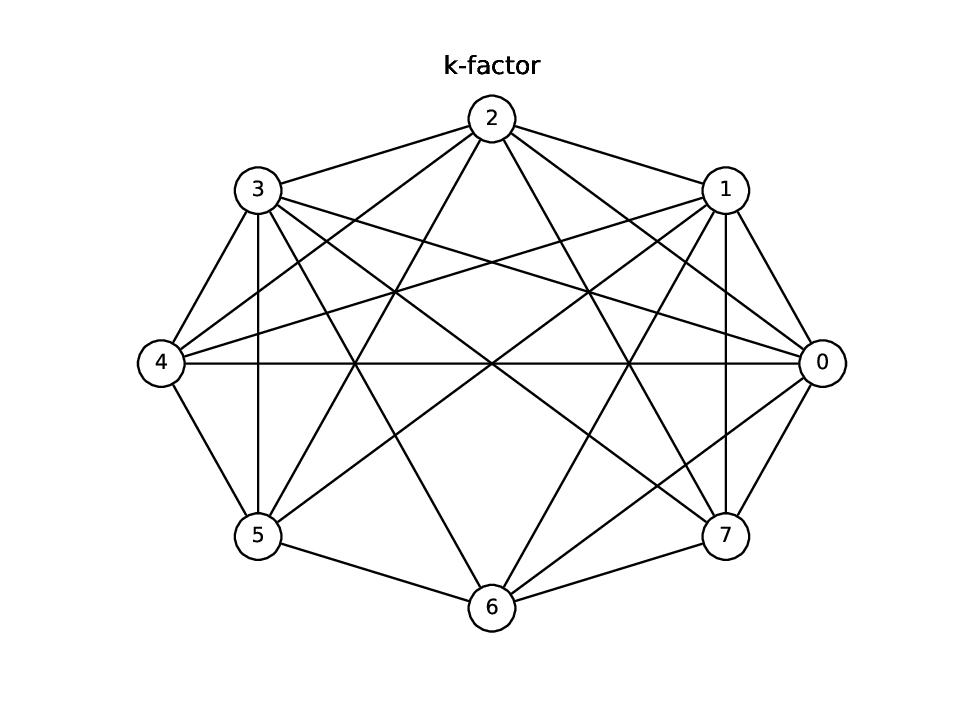}}\hspace{50pt}
	\subfigure{\includegraphics[scale =0.4]{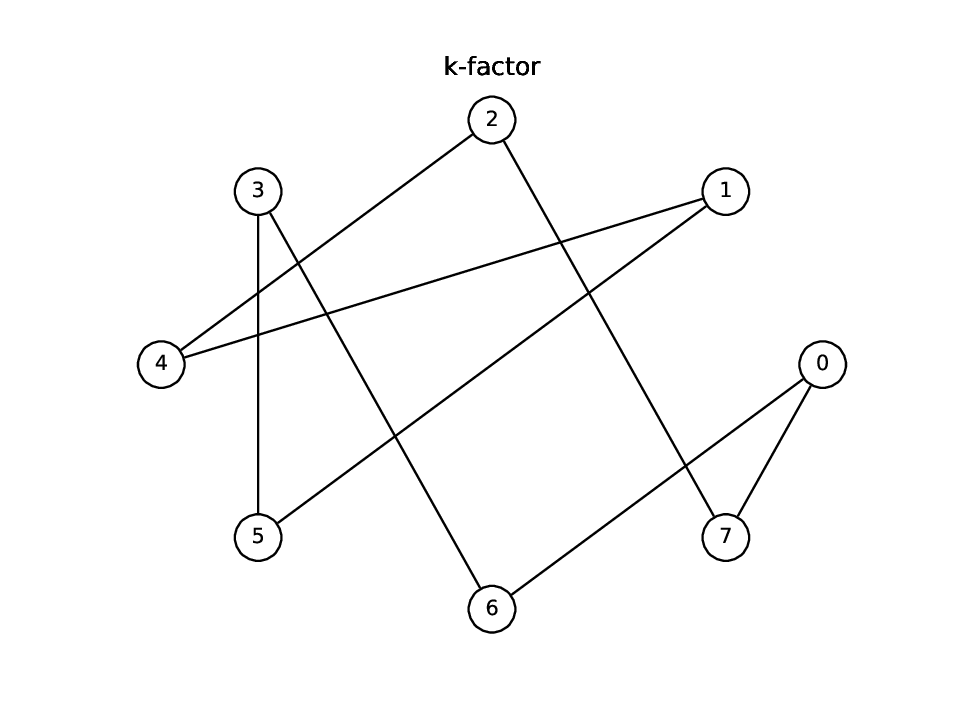}}
	\caption{\em{A graphic realization of 6,6,6,6,5,5,5,5 with a connected 2-factor}}
	\label{fig03}	
\end{figure}
\end{ex}

It remains to address the problem of generating a graphical realization.  
For this, we need to define the notion of \textit{packing}.

\begin{figure}[h!]
	\centering
	\includegraphics[scale =0.6]{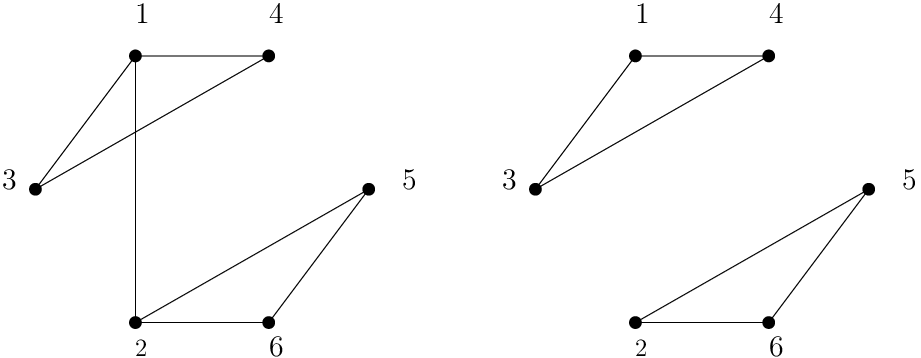}
	\caption{\em{A graph on dSix with a disconnected 2-factor}}
	\label{fig04}
\end{figure}

\begin{definition}
Two graphic sequences $\alpha = (a_1, a_2, \ldots, a_n)$ and $\beta = (b_1, b_2, \ldots, b_n)$ are said to pack if there are graphic realizations $A$ and $B$ of $\alpha$ and $\beta$ respectively on a vertex set $\{v_1, v_2, \ldots, v_n\}$ such that the edges of $A$ and $B$ are disjoint. 
\end{definition}

This means that we have to find a packing of the two sequences $(d_1 -k, d_2-k, \ldots, d_n -k)$ and $(k, k, \ldots, k)$.
When $k=2$, $a=3$ and $b = 2$, this is easy. Construct a 2-factorable graphic sequence that has a prefix sequence made up of an even number of 3's and a suffix sequence of 2's with arbitrary length. The graphic sequence $d -k$ is made up of 1's only and has length equal to the number of 3's. The $k$-factor is 
made up of 2's only and its length is equal to that of the starting sequence. \\

Generate a $2$-regular graph 
$(2, 2, \dots, 2)$ and superpose on this a perfect matching of a sequence  $(1, 1, \ldots, 1)$ with an even number of 1's that is edge-disjoint from 
$2$-regular graph. \\

The algorithm in Section 4 finds a $k$-factor which is a packing but the resulting $k$-factor is not provably connected. It is an interesting problem
to prove that it is always connected or come up with a modification that results in a connected k-factor. \\

Thus for the example graphic sequence $dSix$, we get the graphical realization shown on the left in Fig.~\ref{picdSix} by this packing technique.

\section{Generating $k$-factorable graphic sequences with no connected $k$-factors}

It is an interesting problem to generate examples of $k$-factorable graphic sequences of length $n$ with no connected $k$-factors,  
for $k \geq 2$.\\

Fix an $s < n/2$. Consider sequences of the type: $d = (n-1, \ldots, n-1, x, \ldots x, s, \ldots, s)$ that has a prefix subsequence of length $s$, each of which is $n-1$, 
and a suffix subsequence of length $s$ each of which is $s$. We establish the following claim for this class of sequences for various choices of $k$.\\

From equation~(\ref{raoInequalities}) of Section 2, we get:

\begin{equation*}
(n-1)*s < s(n-s-1) + s*s, 
\end{equation*}

which is false for any such $s$. \\

The following proposition is invoked in all the cases discussed below.

\begin{prop}
	An $r$-regular graph on $n$ vertices can be constructed if $r \leq n-1$ and $nr$ is even. 
\end{prop}

\subsection{Case $k = 2$}
\begin{claim}
Assume $(n-4)*x$ is even. For $4 \leq x \leq n-3$, the sequences $d = (n-1, n-1, x, \ldots x, 2, 2)$ are 2-factorable graphic sequences with no connected 2-factors. 
\end{claim}

{\bf Proof:} We show that both $d$ and $d-2$ are graphic constructively.
To show that $d$ is graphic,  
we saturate the first two vertices of degrees $n-1$ by joining them to each other and to all the remaining vertices. 
This also saturates the last two vertices of degree 2. This leaves us with a susbsequence of length $n-4$, each $x-2$.\\

 We complete by constructing an $x-2$-regular graph on these $n-4$ vertices. \\

A similar argument shows that the sequence $d-2 = (n-3, n-3, x-2, \ldots, x-2)$ is graphic. We construct a graph on $n-2$
vertices, where 
the vertices corresponding to the ones of degree $n-3$ are  each joined to the remaining $n-4$ vertices and to each other. 
This reduces the degree of each of the trailing $n-4$ vertices by 2 and it is easy constuct a regular graph 
on these vertices so that each is of degree $x-4$. $\hfill \Box$ \\

For $d = (15, 15, 6, 6, 6, 6, 6, 6, 6, 6, 6, 6, 6, 6, 2, 2)$, the figures Fig.~\ref{fig05} and Fig.~\ref{fig06} below show graphic realizations of $d$ and $d-2$, 
produced by a program that we wrote. \\

\begin{figure}[h!]
	\centering
	\includegraphics[scale =0.4]{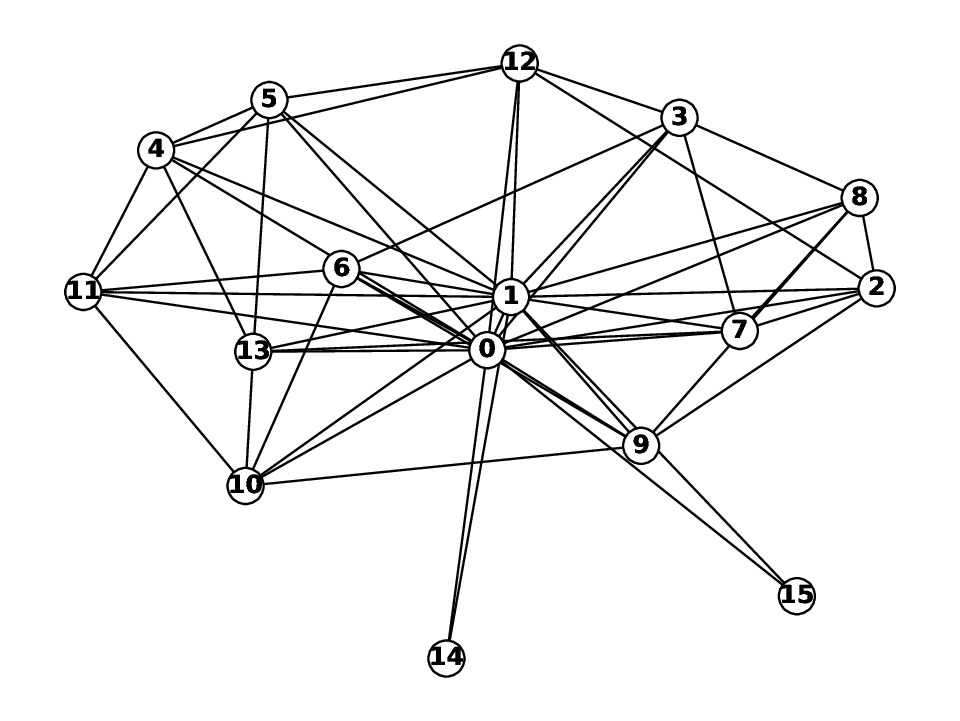}
	\caption{\em{Graph for the degree sequence (15, 15, 6, 6, 6, 6, 6, 6, 6, 6, 6, 6, 6, 6, 2, 2)}}
	\label{fig05}
\end{figure}

\begin{figure}[h!]
	\centering
	\includegraphics[scale =0.4]{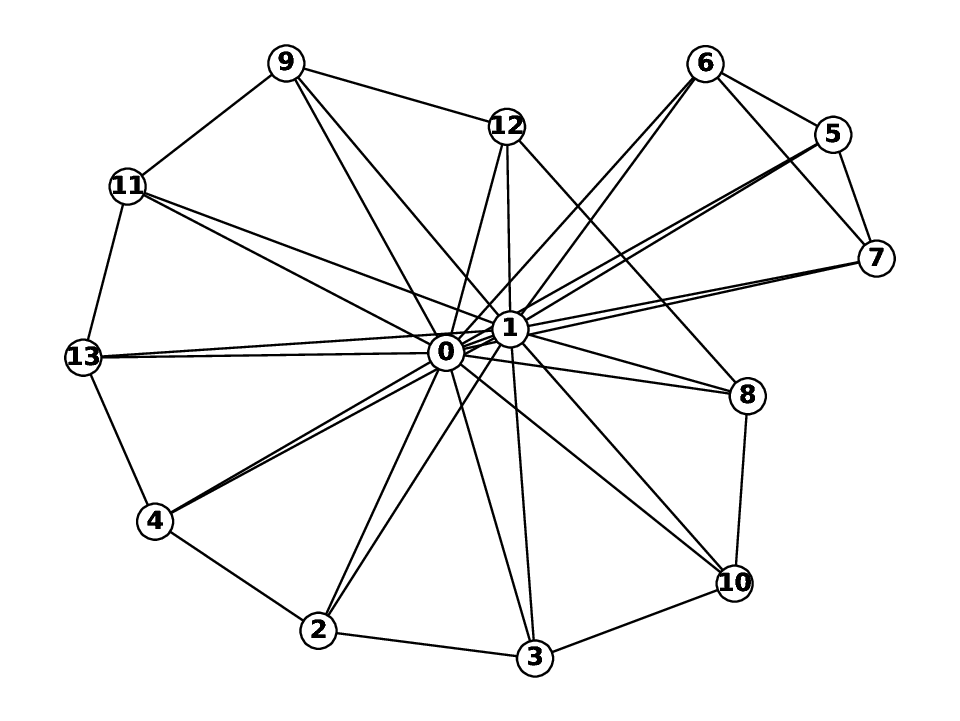}
	\caption{\em{Graph for the degree sequence (13, 13, 4, 4, 4, 4, 4, 4, 4, 4, 4, 4, 4, 4)}}
	\label{fig06}
\end{figure}

\begin{figure}[h!]
	\centering
	\includegraphics[scale =0.4]{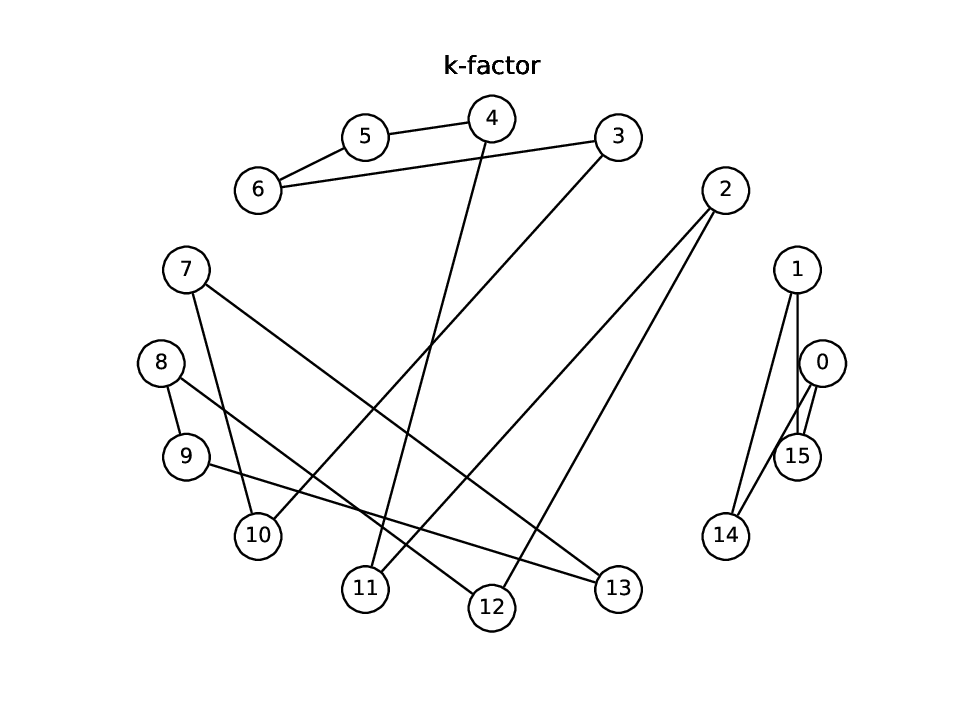}
	\caption{\em{Disconnected 2-factor of the sequence (15, 15, 6, 6, 6, 6, 6, 6, 6, 6, 6, 6, 6, 6, 2, 2)}}
	\label{fig07}
\end{figure}

A disconnected 2-factor is shown in Fig.~\ref{fig07}.\\

Are there other choices of $s$ and the unspecified $n - 2s$ intermediate values for which the sequences are 
graphic, with disconnected $k$ factors for suitable choices of $k$ ?\\

For example, by experimentations, using software we 
have developed, we found that for $n = 10$ and $s = 3$, the sequences $d = (9, 9, 9, 6, 6, 6, 6, 3, 3, 3)$, and $d = (9, 9, 9, 5, 5, 5, 5, 3, 3, 3)$ 
are graphic and have disconnected 2-factors. \\

\begin{figure}[h!]
	\centering
	\includegraphics[scale =0.4]{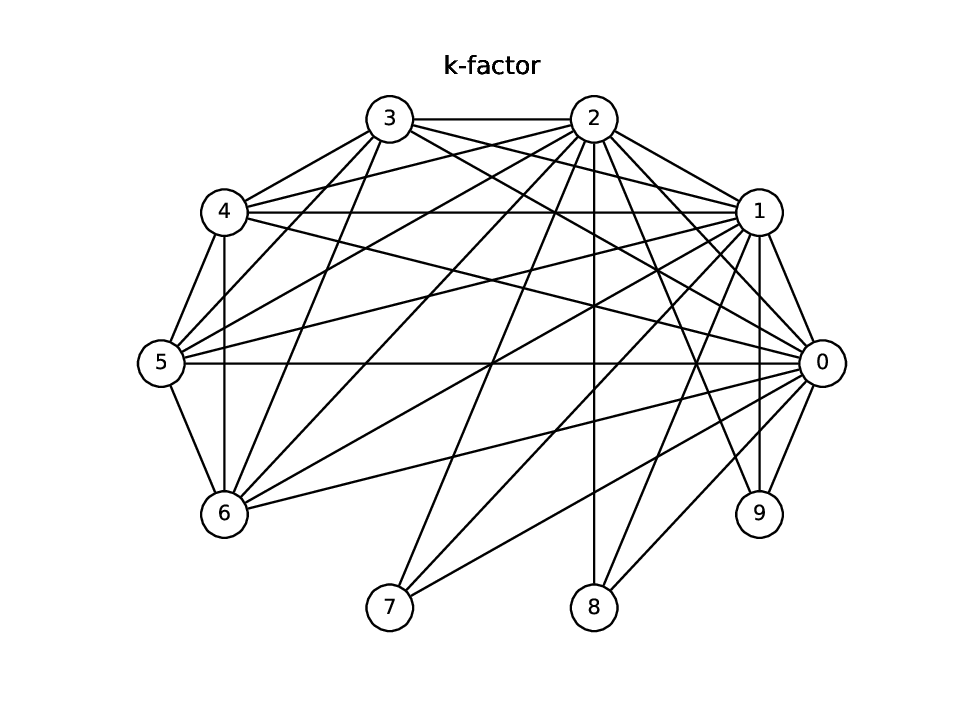}
	\caption{\em{Graph for the degree sequence (9, 9, 9, 6, 6, 6, 6, 6, 3, 3, 3)}}
	\label{fig08}
\end{figure}

\begin{figure}[h!]
	\centering
	\includegraphics[scale =0.4]{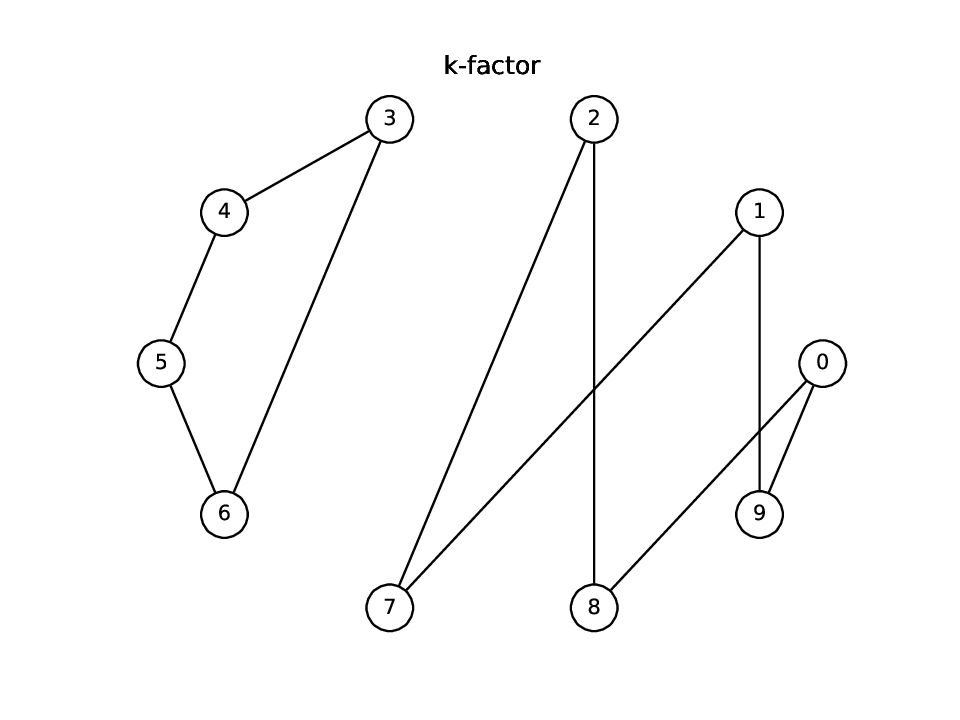}
	\caption{\em{2-factors of the graph for the degree sequence(9, 9, 9, 6, 6, 6, 6, 6, 3, 3, 3)}}
	\label{fig09}
\end{figure}

In Fig.~\ref{fig08} and Fig.~\ref{fig09}, respectively, a realization of the graphic sequence $d = (9, 9, 9, 6, 6, 6, 6, 3, 3, 3)$ and its 2-factor, consisting of two disjoint cylces, are shown.  \\

It is interesting to note that the two disjoint cycles that make up the 2-factor cannot be merged into a 
single cycle. This is because of the structure of the subgraph $C_1$ on the vertices $\{0, 1, 2, 7, 8, 9\}$. Let 
$A_1 = \{7, 8, 9\}$ be the vertices at an even distance from 8, and $A_2 = \{ 0, 1. 2\}$ at an odd distance from 8
in this subgraph. Then the vertices of $A_1$ are independent in the parent graph and the induced subgraph on 
vertices of $A_2$ is complete and each of its vertices is joined to all the vertices of the induced subgraph $C_2$ on the vertices $\{3, 4, 5, 6\}$. The notation $C_1 \rightarrow C_2$ is used to denote this relationship between $C_1$ and $C_2$.\\ 

Rao and Rao ~\cite{RAMACHANDRARAO1972185} proved the following interesting result.

\begin{theorem}
	Let $G$ be a graph with a $k$-factor $F$ consisting of two components $C_1$ and $C_2$. Let $k \geq 2$ and let  $C_1$ and $C_2$ be bicoherent. If the degree
	sequence of $G$ is not connected $k$-factorable then either $C_1 \rightarrow C_2$, or  $C_2 \rightarrow C_1$.
\end{theorem}

\subsection{Case $k = 3$}
\textbf{Claim 1:} Let $n$ to be even and $x$ lie in the range $2*3 \leq x \leq n-4$. Then the sequences 

${d = (n - 1, n - 1, n - 1, x, \ldots, x, 3, 3, 3)}$ are 3-factorable graphic sequences with no connected 3-factors.\\

\textbf{Proof:} We show that both $d$ and $d-3$ are graphic constructively. To show $d$ is graphic, we saturate the first three vertices of degree $n - 1$ by joining them to each other and to all remaining vertices. This also saturates the last three vertices of degree 3. This leaves us with a subsequence of length $n - 6$, each $x - 3$. Construct an $x - 3$ regular graph on these $n - 6$ vertices.
$$(n - 1, n - 1, n - 1, x, \ldots, x, 3, 3, 3) \to (x - 3, \ldots, x - 3)$$

For ${d - 3 = (n - 4, n - 4, n - 4, x - 3, \ldots, x - 3)}$ we construct a graph on $n - 3$ vertices, where the vertices corresponding to the ones of degree $n - 4$ are each joined to the remaining vertices and each other. This reduces the trailing $n - 6$ vertices by 3 and it is easy to construct a regular graph on these vertices so that each of degree $x - 6$.\\
$${(n - 4, n - 4, n - 4, x - 3, \ldots, x - 3) \to (x - 6, \ldots, x - 6)}$$

\subsection{Case $k = s$}
\textbf{Claim 2:} Let $n$ to be even and $x$ lie in the range $2s \leq x \leq n-s-1$. Then the sequences 

${d = (n - 1, n - 1, \ldots, n - 1, x, \ldots, x,s, s, \ldots, s)}$ are $k$-factorable graphic sequences with no connected $k$-factors.\\

\textbf{Proof:} We must show both $d$ and $d - k$ are graphic constructively. To show $d$ is graphic, there are $k$ (or $s$), $n - 1$ degree nodes that we saturate with the remaining nodes and each other. So, the $n - k$ trailing nodes each get saturated $k$ times and the $n - 1$ nodes are saturated $n - 1$ times. There are $k$ (or $s$) trailing $s$ nodes, so these are fully saturated. This leaves us with $(x - k, \ldots, x - k)$ and has $n - 2k$ entries. Complete by constructing an $x - k$ regular graph on these $n - 2k$ vertices.\\
$${(n - 1, \ldots, n - 1, x, \ldots, x, s, \ldots, s) \to (x - k, \ldots, x - k)}$$

Similarly, the sequence ${d - k = (n - k - 1, \ldots, n - k - 1, x - k, \ldots, x - k)}$ is graphic. We construct a graph on $n - k$ vertices. Where the vertices corresponding to the ones of degree $n - k - 1$ are each joined to the remaining $n - 2k$ vertices and each other. This reduces the degree of each trailing $n - 2k$ vertices by $k$. It is easy to construct a regular graph on these vertices of length $n - 2k$ with degree $x - 2k$.\\
$${(n - k - 1, \ldots, n - k - 1, x - k, \ldots, x - k) \to (x - 2k, \ldots, x - 2k)}$$

Here is an algorithm which generates a disconnected $k$-factor of the form $(n-1, ..., n-1,x, ..., x, s, ..., s)$\\

\begin{algorithm}
	\caption{Construction of a random disconnected $k$-factor sequence}
	\begin{algorithmic}[1]
		\State \textbf{Input:} $n$, $k$ : Where $n$ is even, $k = s < n/2$ and $3k + 1\leq n$
		\State \textbf{Output:} A disconnected $k$-factorable graphic sequence
		\Procedure{generateSequenceDisconnected}{$n, k$} 
			\State Set $s = k, k < n/2$
			\State Generate a random number $x$, $2s\leq x\leq n-s-1$
			\For {$i=1$ to $s$}
				\State Append $n-1$ to the sequence
			\EndFor
			\For {$i = 1$ to $n-2s$}
				\State Append $x$ to the sequence
			\EndFor
			\For {$i = 1$ to $s$}
				\State Append $s$ to the sequence
			\EndFor
		\EndProcedure
	\end{algorithmic}
\end{algorithm}

\section{Implementation Issues: $k$-factor computation}

To make the discussion self-contained we address the issue of computing a $k$-factor of a graph. We implemented an algorithm 
based on a very simple and elegant consructive proof of Kundu's result by Chen~\cite{CHEN1988177}. We first explain the algorithm with the help of an example and then describe it formally, following the exposition and notation in Seacrest's thesis \cite{seacrest2011}. \\ 

Consider the graphic degree sequence: $\pi = (3, 3,3, 3, 2, 2)$ so that $n = 6$. Let $k=1$. \\

\begin{figure}
	\centering
	\subfigure{\includegraphics[scale =0.6]{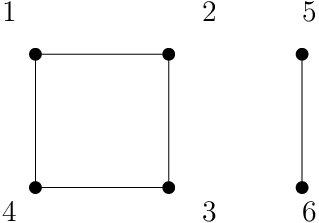}}\hspace{50pt}
	\subfigure{\includegraphics[scale =0.6]{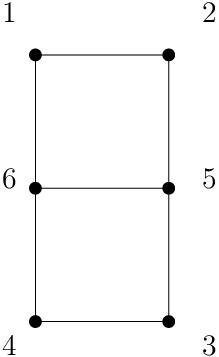}}
	\caption{\em{Graphs A and B}}
	\label{fig10}	
\end{figure}

The graph $A$ of Fig.~\ref{fig10} is an initial realization of the sequence: $\pi - k = \pi - 1= (2,2, 2, 2, 1, 1)$. 
while the graph $B$ of Fig.~\ref{fig10} is an initial realization of the sequence: $n- 1 -\pi = 6 - 1 -\pi= (2,2, 2, 2, 3, 3)$.\\

Superposition of the initial realizations of the graphs $A$ and $B$ is the multigraph $C$ of Fig.~\ref{pic34}.\\

\begin{figure}
	\centering
	\subfigure{\includegraphics[scale =0.6]{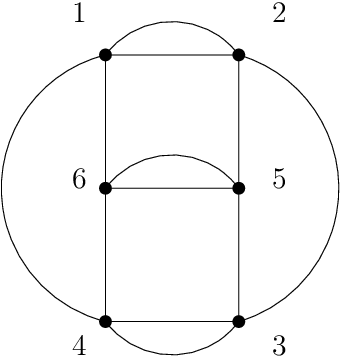}}\hspace{50pt}
	\subfigure{\includegraphics[scale =0.6]{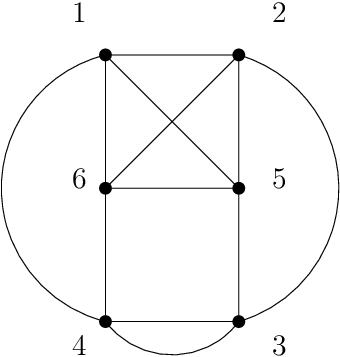}}
	\caption{\em{Graphs C and D}}
	\label{fig11}	
\end{figure}

We remove one of the edges of the multiedge between the vertices 5 and 6 from the graph $B$, by switching the edges $\{1,2\}$ and $\{5,6\}$ in this graph by adding the replacement edges $\{6,2\}$ and $\{5,1\}$ to the same graph. This also removes one of the edges in the multiedge between 1 and 2 and gives us the graph $D$ of Fig.~\ref{fig11}. \\

{\bf Remark:} Why should edge-switching be possible ? Recall that the vertices involved in this edge-switching 
are $u,v, x$ and $y$, where there is no edge between $v$ and $x$, and there is a multiedge between $u$ and $v$. 
Let the total degree of $x$ be expended by joining the set of vertices, say, $V_x$, each of which is connected 
to $x$ by one or two edges. Note that $v$ is not in $V_x$.\\

Now, since $u$ is joined to $v$ by two edges and $x$ is not joined to $v$, the total degree of both $u$ and $x$ being the same, $u$ cannot be joined 
to the vertices in $X$ to match that of $x$. Thus there must a exist vertex $y$ in $X$ such that $x$ is joined to 
$y$ by one edge and there is no edge joining $u$ to $y$, or $x$ is joined to $y$ by two edges, and $u$ is joined to it by
at most one edge. \\

In the first case, edge-switching removes one of the two edges between $u$ and $v$. In the second case, one of the edges
between each of the pairs $\{u, v\}$ and $\{{x, y}\}$ is removed and possibly an edge is introduced bweteen $u$ and $y$ 
to make it a multiedge. In each case, however, there is a net reduction of the total number of multiedges. \\

Consider the graph $C$ of Fig.~\ref{fig11}. Let $u = 6$ and $v = 5$. Since the total degree of $v$ is 4 with degrees
used up bu the edges from the vertex $u$ into it, the remaining 2 degrees are spread among vertices 1, 2, 3 and 4. 
Thus there exists a vertex it is not jouned to. In this case these are vertices 1 and 4. Set $x = 1$. The total degree
4 of 1 is distributed among the vertices $\{6, 2, 4\}$. This is $V_x$. Note that $v = 5$ is not in $V_x$. Since $u = 6$ 
is joined to $v = 5$ with 2 edges and is also joined to 1, it has only 1 more degree to expend on the vertices 2 and 4 
in $V_x - \{6\}$. On the other hand, 1 has three more degrees to expend on the vertices 2 and 4. Hence there exists 
a vertex in  $V_x - \{6\}$ that 6 is not connected to. In this case it is 2. $\Box$\\

\begin{figure}[h!]
	\centering
	\includegraphics[scale =0.6]{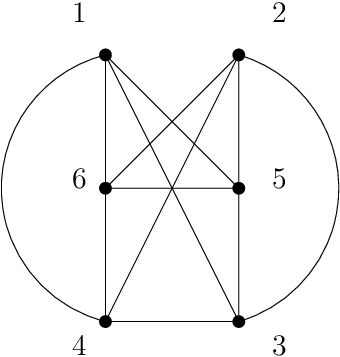}
	\caption{\em{Graph E}}
	\label{fig12}
\end{figure}

We still have a multiedge between 3 and 4 in the superposed graph. We switch the pair of edges $\{1,2\}$ and 
$\{3,4\}$ in the graph $A$ with the replacement pair $\{1,3\}$ and $\{2,4\}$. This gives us the graph $E$  of Fig.~\ref{fig12}.\\

\begin{figure}[h!]
	\centering
	\includegraphics[scale =0.6]{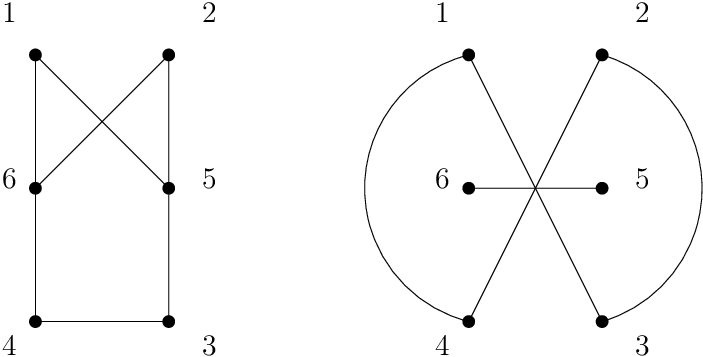}
	\caption{\em{New Graphs B and A}}
	\label{fig13}
\end{figure}

The new realizations of the graphs $B$ (figure on left) and $B$ (figure on right) are now shown in Fig.~\ref{fig13}. \\

\begin{figure}[htbp!]
	\centering
	\includegraphics[scale =0.6]{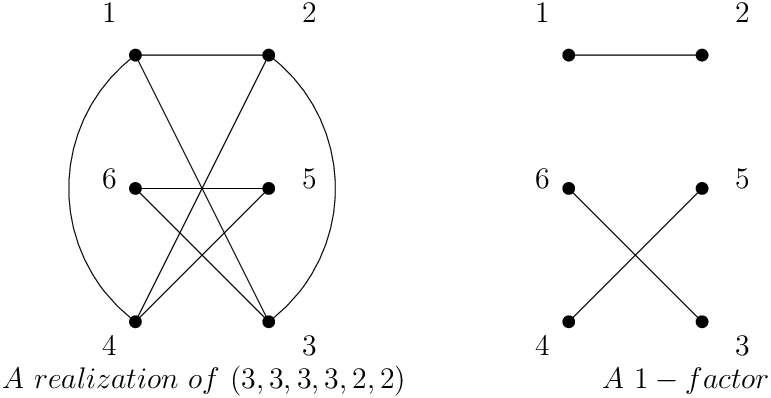}
	\caption{\em{A realization of (3, 3, 3, 3, 2, 2) and a 1-factor of this}}
	\label{fig14}
\end{figure}

The complement of the new graph $B$ is a realization of $\pi = (3, 3,3, 3, 2, 2)$. Both this graph (figure on left) and a 
$1$-factor (figure on right), which is the graph $\overline{B}\backslash A$, are shown in Fig.~\ref{fig14}. \\

\newpage
We give a formal description of the algorithm below.\\

\begin{algorithm}
	\caption{Construction of a $k$-factor}
	\begin{algorithmic}[1]
		\State \textbf{Input:} A $k$-factorable graphic sequence, $d$
		\State \textbf{Output:} A $k$-factor of a graph with degree sequence $d$
		\Procedure{$k$-factor}{$d, k$}
		\State  Compute a realization $A$ of the graphic sequence $(d_1-k, d_2-k, \ldots, d_n-k)$.
		\State  Compute a realization  $B$ of the graphic sequence $(n-1-d_1, n-1-d_2, \ldots, n-1-d_n)$.
		\State  Superpose $A$ and $B$ and make a list $L$ of the multiple edges in the graph $A \cup B$.
		\While {there is a multiedge $\{u, v\}$ in $A \cup B$} 
		     \State  Find a vertex $x$ that is not joined to $v$ and a vertex $w$ that is connected to $u$ by at most one edge.
		      \State Switch edges $\{u, v\}$ and $\{x, y\}$ with new edges 
		      $\{v, x\}$ and $\{u, w\}$ in one of the graphs $A$ or $B$.
		\EndWhile 
		\EndProcedure
	\end{algorithmic}
\end{algorithm}

We have implemented the above algorithm in Python. 

\section{Complexity Analysis}
Though an algorithm is implied by Theorem 16 of Seacrest's thesis, an analysis of the algorithm is missing. Since such an analysis could be useful for coming up with a more efficient algorithm, we add this discussion here. \\

We first have to consider the complexity of finding overlapping edges in the superpositons of the graphs $A$ and $B$. 
This can be done by taking a logical AND of the entries of the adjacency matrices of $A$ and $B$ and then scanning the 
entries of the resutling matrix for 1's. \\

The next step is to consider the complexity of removing multiple edges. For each multiedge $\{u, v\}$, we have find 
vertices $x$ and $y$ to perform edge switching in one of the graphs $A$ or $B$. Finding vertices $x$ and $y$ takes $O(n)$
time, while edge-switching can be done in $O(1)$ time.\\

After each multiple edge removal we might have to update the list of multiple edges as we might remove two multiple 
edges and add a new one. \\

From this analysis we conclude that the complexity of Chen's algoritrhm is $O(n^2 + mn)$, where $m$ is the 
initial number of multiple edges found by the logical AND operation of the adjacency lists of the graphs $A$ and $B$. 

\section{A web application for visualizing $k$-factors}
A Web Application is available to explore at \href{https://github.com/jere-mie/graph-webapp}{https://github.com/jere-mie/graph-webapp}.
Follow the README.md instructions to run the application.

\subsection{Walkthrough}

On the webapps homepage click on the blue "Start Graphing" button. Under the "K-Factors" label click the "open" button.\\

Here you may input a k-factorable graphic sequence and the value of k in the text boxes. When you press "Run" it will show the realizations of the orginal, $d - k$, and $k$-factor graphs.\\ 

The two buttons labeled "Connected Sequence" and Disconnected Sequence" will populate the input boxes with random connected or disconnected $k$-factorable graphic sequences. The connected sequences generated have a connected $k$-factorable realization, but the realization shown when "Run" is clicked may or may not give a connected realization. The disconnected sequences generated will always display disconnected realizations when run.\\

\begin{figure}[h!]
	\centering
	\includegraphics[scale =0.4]{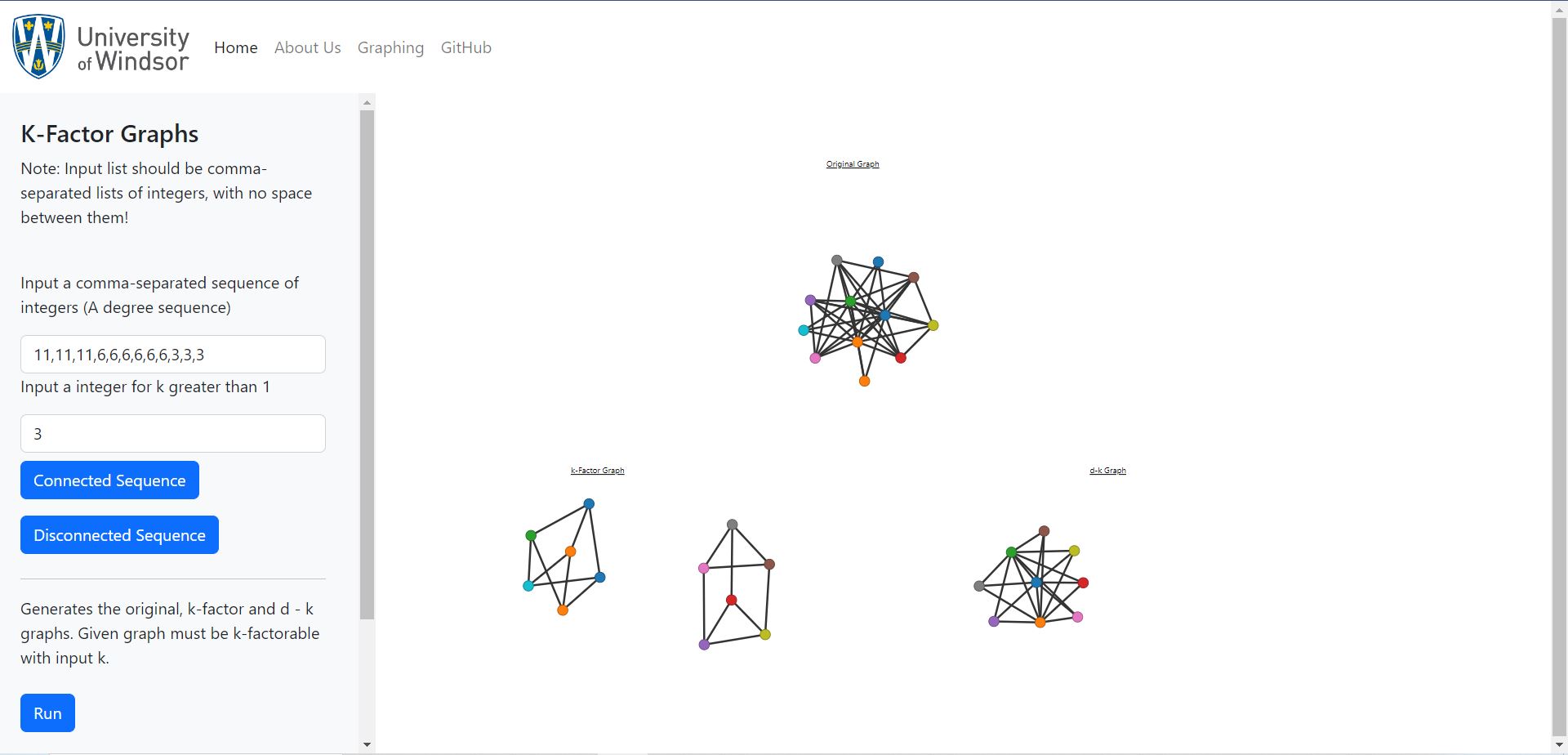}
	\caption{\em{Connected $k$-factor webapp example}}
	\label{fig15}
\end{figure}

\begin{figure}[h!]
	\centering
	\includegraphics[scale =0.4]{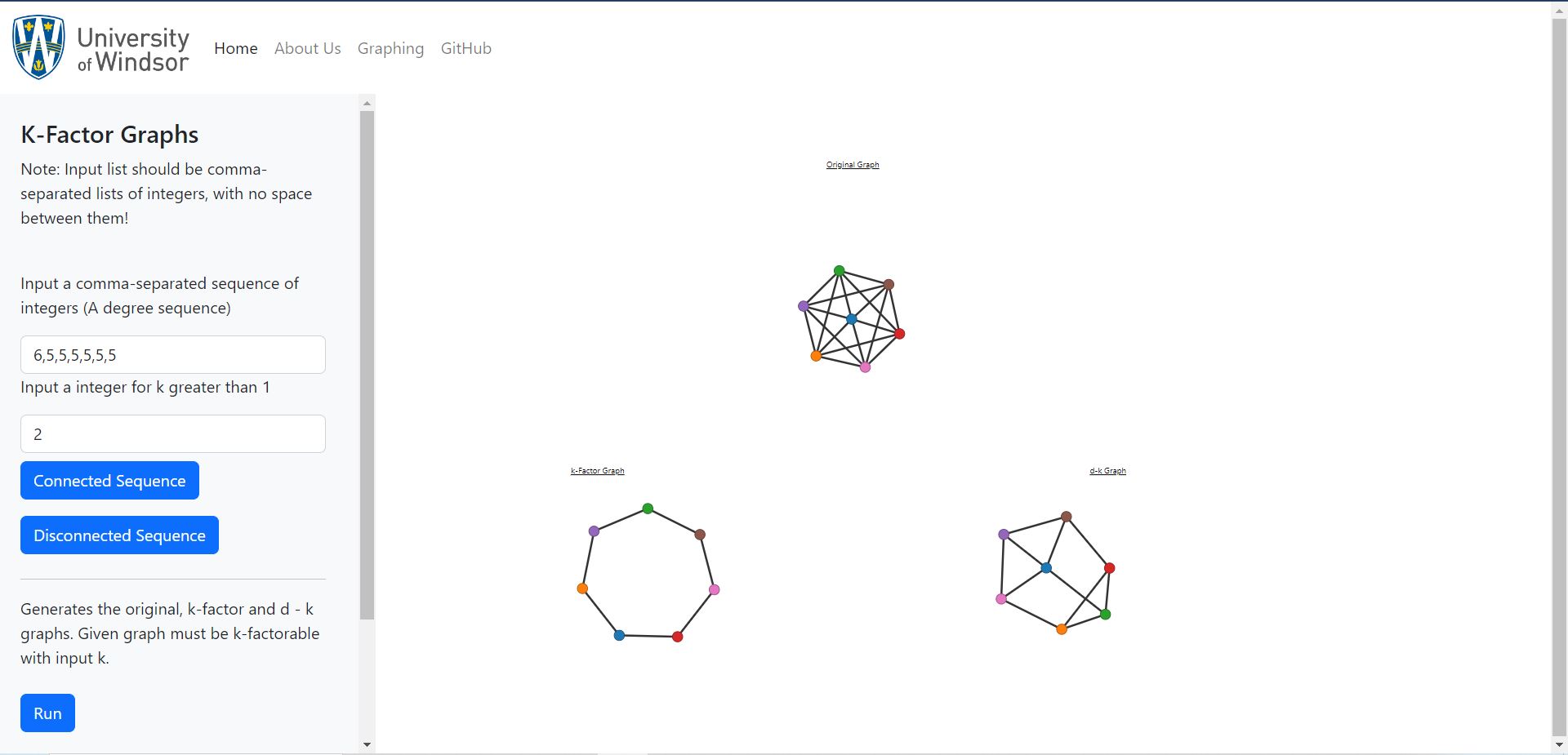}
	\caption{\em{Disconnected $k$-factor webapp example}}
	\label{fig16}
\end{figure}

\clearpage

\section{Conclusions}
In this note, we have considered the previously unexplored problem of generating $k$-factorable graphic sequences that have connected $k$-factors and 
those that have disconnected $k$-factors. Such generation is useful by way of providing us with instances of such graphs that can be used as test cases 
and validating conjectures about such graphs. \\

It might be interesting to come up with other
classes of $k$-factorable graphic sequences with disconnected $k$-factors. We have just scratched the surface of this problem.\\

It might be also useful to find other applications of $k$-factorable graphs, besides the one in Computer Graphics \cite{DBLP:journals/vc/Diaz-GutierrezG05}.

\bibliography{referencesFDS}
\bibliographystyle{abbrv}  

\end{document}